\documentclass[reprint,aps,twocolumn,pra,groupedaddress,10pt]{revtex4}
\usepackage{hyperref}
\usepackage[latin1]{inputenc}
\usepackage{amsmath}
\usepackage{amssymb}
\usepackage{amsthm}
\usepackage{fontenc}
\usepackage{graphicx, array, exscale, dsfont, amsfonts, verbatim, fancyhdr, natbib, bbm,mathrsfs,textcomp,listings, epsfig}

\newcommand{\ket}[1]{\left\vert{#1}\right\rangle}

\newcommand{\epsbar}{\bar{\varepsilon}}

\newcommand{\proj}[1]{\left\vert{#1}\right\rangle \left\langle{#1}\right\vert}

\DeclareMathOperator{\tr}{tr}

\newcommand{\dist}[1]{\frac{1}{2}\left|\left|#1\right|\right|_1}

\newcommand{\ketbra}[2]{{|#1\rangle\!\langle#2|}}
\newcommand{\id}{{1\!\!1}}
\newcommand{\eps}[1]{\varepsilon_{\mathrm{#1}}}

\begin{document}

\newtheorem{thm}{Theorem}
\newtheorem{cor}{Corollary}
\newtheorem{lem}{Lemma}
\newtheorem{defi}{Definition}
\newtheorem{ex}{Example}
\newtheorem*{note}{Note}
\newtheorem*{rem}{Remark}
\newcommand{\qudit}[1]{\left\vert #1 \right\rangle}
\newcommand{\rqudit}[1]{\left\langle #1 \right\vert}
\newcommand{\outerp}[1]{\left\langle #1 \vert #1 \right\rangle}
\newcommand{\Z}{\mathbb{Z}}
\newcommand{\oxc}[1]{{\mathbb{\tiny#1}}}
\newcommand{\F}{\mathbb{F}}
\newcommand{\R}{\mathbb{R}}
\newcommand{\supp}{\mathrm{supp}}
\newcommand{\C}{\mathbb{C}}
\newcommand{\1}{\mathbb{1}}
\newcommand{\cB}{\mathcal{B}}
\newcommand{\cE}{\mathcal{E}}
\newcommand{\cF}{\mathcal{F}}
\newcommand{\cH}{\mathcal{H}}
\newcommand{\cM}{\mathcal{M}}
\newcommand{\cN}{\mathcal{N}}
\newcommand{\cP}{\mathcal{P}}
\newcommand{\cS}{\mathcal{S}}
\newcommand{\cX}{\mathcal{X}}
\newcommand{\cY}{\mathcal{Y}}
\newcommand{\cZ}{\mathcal{Z}}
\newcommand{\CC}{\mathbb{{\mathit{C}}}}
\newcommand{\y}[1]{\tiny{$Y_{#1}$}}
\newcommand{\x}[1]{\tiny{$X_{#1}$}}
\newcommand{\z}[1]{\tiny{$Z_{#1}$}}
\newcommand{\bn}{\mathbf{n}}
\newcommand{\bQ}{\mathbf{Q}}
\newcommand{\bP}{\mathbf{P}}
\newcommand{\blam}{\boldsymbol{\lambda}}
\newcommand{\pr}[1]{{\ketbra{#1}{#1}}}
\newcommand{\ball}[2]{{\cal B}^{#1}\left(#2\right)}
\newcommand{\trz}[2]{\mathrm{tr}_{#1}\left( #2\right)}
\newcommand{\epsbarec}{\bar{\varepsilon}_{\mathrm{EC}}}
\newcommand{\ren}[3]{S_{#1}^{#2}\left(#3\right)}
\newcommand{\hmin}[3]{H_{\mathrm{min}}^{#1}\left(#2|#3\right)}
\newcommand{\hnull}[3]{H_{0}^{#1}\left(#2|#3\right)}
\newcommand{\hmax}[3]{H_{\mathrm{max,old}}^{#1}\left(#2|#3\right)}
\newcommand{\hmaxnew}[3]{H_{\mathrm{max,new}}^{#1}\left(#2|#3\right)}
\newcommand{\leak}{\mathrm{leak}_{\mathrm{EC}}} 
\newcommand{\ent}{Ent\left(\epsbar,\rho_{X^nE^n}\right)}
\newcommand{\xia}{\xi_{\mathrm{att}}}

\title{Secret key rates for coherent attacks}
\date{\today}
\author{Markus Mertz}
\email{mertz@thphy.uni-duesseldorf.de}
\author{Hermann Kampermann}
\author{Sylvia Bratzik}
\author{Dagmar Bru{\ss}}
\affiliation{Institute for Theoretical Physics III, Heinrich-Heine-Universit\"at D\"usseldorf, 40225 D\"usseldorf, Germany.}

\begin{abstract}
We develop a new method to quantify the secret key rate for permutation-invariant protocols for coherent attacks and finite resources. The method reduces the calculation of secret key rates for coherent attacks to the calculation for collective attacks by bounding the smooth min-entropy of permutation-invariant states via the smooth min-entropy of corresponding tensor-product states. The comparison of the 
results to the well-known post-selection technique for the BB$84$ and six-state protocol shows the high relevance of this method. 
Since our calculation of secret key rates for coherent attacks strongly depends on the way of treating collective attacks, a prospective progress in the analysis of collective attacks will immediately cause progress in our strategy.
\end{abstract}
\maketitle

\section{Introduction}
The aim of quantum key distribution (QKD) is the generation of a secret key between two authorized parties Alice and Bob in the presence of an eavesdropper Eve. In practical implementations the number of signals used to establish a secure key is finite. An essential element of the calculation of key rates for a finite number of signals is
the evaluation of the smooth min-entropy \cite{RenPhD} for high-dimensional states, which is in general hard or even impossible to compute. 
In the last years many results have appeared \cite{RenPhD,Mey06,Sca08,Sca08a,She10,Cai09,She10a,Bra11,Abr11,Tom11,Hay11} considering the calculation of secret key rates for finite resources under the restriction of the eavesdropper's attack to a collective attack \cite{Biham2002,Bih97}, where Eve interacts with each signal independently and identically. This restriction leads to a state, which has tensor-product form and allows to bound the smooth min-entropy by the conditional von Neumann entropy of a single-signal state by using the asymptotic equipartition property (AEP) \cite{RenPhD,Tom09}. 

In studies of coherent attacks \cite{Cir97,Bech99} the eavesdropper is not restricted at all, i.e. she may interact with all signals simultaneously. Already in the year $2005$ it was shown in \cite{Kra05,Ren05a} that for protocols, which are invariant under permutations of single-signal states, collective and coherent attacks are equivalent in the case of infinitely many signals. But for a finite number of signals this equivalence has not been proven yet. As a consequence the development of tools to compute a secret key for finite resources in the presence of coherent attacks is necessary.

Up to now direct strategies that treat coherent attacks only exist for the BB$84$ \cite{BB84} protocol (see \cite{Tom11},\cite{Hay11}). In \cite{Tom11} Tomamichel et al used an uncertainty relation for smooth entropies \cite{Tom11a} to circumvent the evaluation of the smooth min-entropy by the computation of the smooth max-entropy \cite{RenPhD}. Since the resulting max-entropy has to be evaluated for a classical state, the calculation becomes analytically solvable.

In comparison to these direct strategies, many studies have focused on indirect approaches like post-selection \cite{Chri09} or the de Finetti approach \cite{RenPhD,Ren07} to quantify secret key rates, where the analysis for coherent attacks is traced back to the investigation of collective attacks. In \cite{She10a}, these indirect approaches have been compared to each other for the BB$84$ protocol with the result, that the post-selection technique exceeds the de Finetti approach in terms of secure key rates.

In this paper we present a new strategy to calculate secret key rates for general permutation-invariant (i.e. the output of the protocol remains the same under permutations of the input pairs) protocols for coherent attacks. In particular, we relate the secret key rate for coherent attacks to the calculation of secret key rates for collective attacks by bounding the smooth min-entropy of a permutation-invariant state via the min-entropy of a corresponding tensor-product state ``smoothed'' over a reduced environment. We compare the results to the post-selection technique by applying the AEP-bound for the treatment of collective attacks. Note that most of the protocols studied in the literature already fulfill the condition of permutation-invariance or can made to be permutation-invariant, like e.g the BB$84$ and six-state \cite{Bru98,Bec99} protocol.

The paper is organized as follows. 
In Section~\ref{SEC:Preliminaries} we explain the protocol and fix the notation. We clarify the formalism used to calculate secret key rates under the assumption of collective attacks in Section~\ref{SEC:Collective}. The formalism to analyze coherent attacks, the main result of this paper, is presented in Section~\ref{SEC:Coherent}. Section~\ref{SEC:PostSelection} shortly reviews the post-selection technique, which is then compared to the new strategy with respect to secret key rates for the BB$84$ and six-state protocol in Section~\ref{SEC:Comparison}. Finally, Section~\ref{SEC:Conclusion} concludes the paper.

\section{\label{SEC:Preliminaries}Preliminaries}
In this paper we consider permutation-invariant entanglement-based QKD protocols, which consist of the steps: state distribution, sifting, parameter estimation (PE), error correction (EC), error verification and privacy amplification (PA) (for a detailed description see \cite{Kra05,Ren05a}). Here, permutational invariance means that for any permutation of the input pairs the output of the protocol remains unchanged.
In the following we denote by $\rho_{AB}^N$ the initial state of $N$ signals shared by Alice and Bob, and by $\rho_{ABE}^N$ a purification of $\rho_{AB}^N$, which describes the state shared by Alice, Bob and Eve after the state distribution. Now, let $\cN_{AB}$ be the operation, that represents the procedures, which Alice and Bob perform on their states, i.e. measurement, sifting, parameter estimation, error correction and error verification. (Note that privacy amplification is not included here, since the output of this procedure is the final bit-string used as key.)
Then we define the resulting classical-quantum state containing Alice's bit string and Eve's quantum state as $\rho_{XE}^n:=\left(\mathcal{N}_{AB}\otimes \id_E \right)\rho_{ABE}^N$. As the main quantity for the calculation of secret key rates we use the smooth min-entropy \cite{RenPhD} 
\begin{equation}
  \hmin{\eps{}}{\rho_{AE}}{E}:= \sup_{\sigma_{AE} \in \ball{\frac{\eps{}}{2}}{\rho_{AE}}} \sup_{\rho_E \in \cS(\mathcal{H}_E)} \hmin{}{\sigma_{AE}}{\rho_E},
\end{equation}
defined as an optimization of the min-entropy
\begin{equation}
   \hmin{}{\sigma_{AE}}{\rho_E}:=\sup{\left\lbrace \lambda \in \mathbb{R}:2^{-\lambda}\id_A\otimes \rho_E-\sigma_{AE}\geq0\right\rbrace }
 \end{equation} 
over an $\frac{\eps{}}{2}$-environment given by 
\begin{equation}
  \ball{\frac{\eps{}}{2}}{\rho}:=\left\lbrace \sigma: \dist{\sigma -\rho}\leq \frac{\eps{}}{2} \right\rbrace,
\end{equation}
with the $1$-norm $\left|\left|A\right|\right|_1=\tr\left(\sqrt{AA^\dagger}\right)$. $\cS(\mathcal{H}_E)$ denotes the set of density operators on the Hilbert space $\mathcal{H}_E$.

\section{\label{SEC:Collective}Collective attack}
In contrast to coherent attacks, the assumption of collective attacks forces the eavesdropper Eve to interact with each of the signals separately. Under this restriction the distributed state can for permutation-invariant protocols be regarded as a product state $\rho_{AB}^{\otimes N}$, which is diagonal in the Bell-basis \cite{Kra05,Ren05a}. We denote by $m$ the number of randomly chosen signals used for parameter estimation and by $n$ the remaining number of signals for privacy amplification. Then, the rate of an $\eps{}$-secure key can be quantified in the following way. 

\begin{thm}\cite{Sca08} Let $\eps{PE}, \eps{EC}, \eps{PA}, \epsbar > 0$ and let $\rho_{XE}^{\otimes n}=\left(\cN_{AB}\otimes \id_E \right)\rho_{ABE}^{\otimes N}$ be a tensor-product state for a purification $\rho_{ABE}$ in $\cH_{ABE}$ of the state $\rho_{AB}\in \cS(\cH_{AB})$. Then the rate of an $\eps{coll}:=(\eps{PE}+\eps{EC}+\eps{PA}+\epsbar)$-secure key is given by 
\begin{equation}\label{eq:PACOLL}
 r:=\frac{1}{N}\inf_{\rho_{AB}\in \Gamma_{\mathrm{coll}}}\left(\hmin{\epsbar}{\rho_{XE}^{\otimes n}}{E}-\leak \right)+\frac{2}{N}\log_2\left(2\eps{PA}\right).
\end{equation} 
\end{thm}
The smooth min-entropy of the classical-quantum state $\rho_{XE}^{\otimes n}$ shared by Alice and Eve and the correction $2\log_2\left(2\eps{PA}\right)$ arise from the analysis of privacy amplification. The entropy quantifies Eve's uncertainty of Alice's bit-string. 

The term $\leak$ stands for the number of bits which Alice and Bob leak to the eavesdropper due to public communication during the error correction procedure and cost for the error verification. In total, the leakage can be estimated by \cite{Sca08,Tom11}
\begin{equation}
\leak :=n 1.1 H(X|Y)+\log_2{\left(\frac{2}{\eps{EC}}\right)}.
\end{equation} Here, the factor $1.1$ denotes the efficiency of a specific error-correction protocol used during the key-generation.
The minimization of the smooth min-entropy is due to parameter estimation, where we only except qubit-states $\rho_{AB}$ which are contained in the set \cite{Tom11} 
\begin{equation}\label{eq:PEcoll}
\Gamma_{\mathrm{coll}}:=\left\lbrace \sigma_{AB}
 : \dist{P_m-P_n} \leq \xi\left(\eps{PE},n,m\right)\right\rbrace
\end{equation} with 
\begin{equation}\label{eq:PEcoll1}
\xi\left(\eps{PE},n,m\right):=\sqrt{\frac{(n+m)(m+1)\ln{\left(1/\eps{PE}\right)}}{8m^2n}}.
\end{equation}
This means, that the tolerated quantum bit error rate ($QBER$) $P_m$ due to an $m$-fold independent application of a $POVM$ $\cE$ on a tensor-product state 
is $\xi$-close to the parameter $P_n$, which corresponds to a virtual measurement on the remaining $n$ signals, which are used for the key generation, except with probability $\eps{PE}$ (see Lemma~\ref{lem:tom} in the Appendix). Note that this estimate has been developed in \cite{Tom11} for coherent attacks, i.e. Lemma~\ref{lem:tom} holds for permutation-invariant states. As tensor-product states in collective attacks are permutation-invariant, Lemma~\ref{lem:tom} can be applied.

For product states $\rho_{XE}^{\otimes n}$ we can use the asymptotic equipartition property (see Eq.~(\ref{eq:minprod})) to bound the smooth min-entropy by the conditional von Neumann entropy of a single copy $\rho_{XE}$.
Finally, we get for the rate 
of an $\eps{coll}:=(\eps{PE}+\eps{EC}+\eps{PA}+\epsbar)$-secure key: 
 \begin{eqnarray}\label{eq:ratecoll}
 r_{\mathrm{coll}}&:=&\frac{n}{N}\Bigg[\inf_{\rho_{AB}\in\Gamma_{\mathrm{coll}}}\left( S(X|E)-\frac{\leak}{n} \right) \nonumber \\
                  &-& 5\sqrt{\frac{\log_2(2/\epsbar)}{n}}\Bigg]+\frac{2}{N}\log_2\left(2\eps{PA}\right)
 \end{eqnarray} where 
\begin{equation} 
S(X|E)=S(\rho_{XE})-S(\rho_E)
\end{equation} with $S(\rho):=-\tr{\left(\rho \log_2{\rho}\right)}$.

In the next section we present a formalism to treat coherent attacks. We will see that the analysis of secret key rates for coherent attacks can be traced back to the calculation of secret key rates under the assumption of collective attacks (see Eq.~(\ref{eq:ratecoll})).

\section{\label{SEC:Coherent}Coherent attack}
A coherent attack is the most general attack an eavesdropper can perform, i.e. Eve is not restricted at all. For the investigation of secret key rates for coherent attacks, we have to consider non-product states for the evaluation of the smooth-min entropy. No changes are needed in the analysis of parameter estimation for collective attacks (see Eq.~(\ref{eq:PEcoll})), because it also holds for coherent attacks (i.e. non-product states (see Lemma~\ref{lem:tom} in the Appendix)). Since error correction and error verification are also independent of the underlying attack of the eavesdropper (they are purely classical procedures), the protocol analysis for these steps can be adopted from the one for collective attacks. 

For permutation-invariant protocols it has been shown in \cite{Kra05} and \cite{Ren05a} that we can assume w.l.o.g. that, after the distribution of $N$ qubit pairs, Alice and Bob share a permutation-invariant quantum state, which is a convex combination of tensor-products of Bell-states:
\begin{equation}\label{eq:permu}
\rho_{AB}^{N}=\cP_N\left(\sum_{\bn \in \Lambda^N} \mu_{\bn}\sigma_1^{\otimes n_1}\otimes \sigma_2^{\otimes n_2}\otimes \sigma_3^{\otimes n_3}\otimes \sigma_4^{\otimes n_4}\right)
\end{equation} with probabilities $\mu_{\bn}$ for the ``realization'' $\bn$ and the set of realizations 
\begin{equation}
\Lambda^N:=\left\lbrace \bn=\left(n_1,n_2,n_3,n_4\right):\sum_{i=1}^4 n_i =N \right\rbrace.
\end{equation} The $\sigma_i$ for $i=1,..,4$ correspond to the projector onto the $4$ Bell-states in $\cH_A \otimes \cH_B$, i.e.
\begin{eqnarray}
 \sigma_1 &=& \proj{\phi^+}, \nonumber \\
 \sigma_2 &=& \proj{\phi^-}, \nonumber \\
 \sigma_3 &=& \proj{\psi^+}, \nonumber \\
 \sigma_4 &=& \proj{\psi^-}, 
\end{eqnarray} with
\begin{eqnarray} 
\ket{\phi^\pm}&:=&\frac{1}{\sqrt{2}}\left(\ket{00}\pm \ket{11}\right) \qquad  \mathrm{and} \\
\ket{\psi^\pm}&:=&\frac{1}{\sqrt{2}}\left(\ket{01}\pm \ket{10}\right). 
\end{eqnarray}
$\cP_{N}$ denotes the completely positive map (CPM) which symmetrizes the state with respect to all possible distinguishable permutations of the $N$ qubit pairs.

The following section explains the analysis of parameter estimation for permutation-invariant states (see Eq.~(\ref{eq:permu})).

\subsection{Parameter estimation}\label{SUBSEC:PE}
Let the sifting procedure now be such that $N_s=n+m$ signals remain, where $m$ denotes the number of randomly chosen signals used for parameter estimation and $n$ denotes the remaining number of signals for privacy amplification. Then we can adopt Lemma~\ref{lem:tom} to estimate the $QBER$ $Q_n$ by the tolerated $QBER$ $Q_m$ coming from a measurement on general permutation-invariant states (see also the arguments below Eq.~(\ref{eq:PEcoll1})). 
\begin{thm}Let $\eps{PE}>0$ and $m+n=N_s$. Let $\rho_{AB}^{N_s} \in \cS\left(\cH_{AB}^{\otimes {N_s}}\right)$ be a permutation-invariant quantum state, and let $\cE$ be a $POVM$ on $\cH_{AB}$ which measures the $QBER$. Let $\bQ_m$ and $\bQ_n$ be the frequency distributions when applying the measurement $\cE^{\otimes m}$ and $\cE^{\otimes n}$, respectively, to different subsystems of $\rho_{AB}^{N_s}$. Then for any element $Q_m$ and $Q_n$ from $\bQ_m$ and $\bQ_n$ except with probability $\eps{PE}$
\begin{equation}\label{eq.PEpermu}
\dist{Q_m-Q_n}\leq \xi\left(\eps{PE},n,m\right)
\end{equation} 
with $\xi(\eps{PE},n,m):=\sqrt{\frac{(m+n)(m+1)\ln{\left(1/\eps{PE}\right)}}{8m^2n}}$.
\end{thm}
\emph{Proof:} This follows directly from Lemma~\ref{lem:tom} in the Appendix, which is a consequence of \cite{Tom11}.\qed

Now with the definition of the set of states, which pass the parameter estimation procedure
 \begin{equation}
\Gamma^n_{\eps{PE}}:=\left\lbrace \sigma^n_{AB}
 : \dist{Q_m-Q_n} \leq \xi\left(\eps{PE},n,m\right)\right\rbrace, 
 \end{equation} we are able to give an analytic expression for the rate of an $\eps{}$-secure key for coherent attacks. 

\begin{cor}Let $\eps{PE}, \eps{EC}, \eps{PA}, \epsbar > 0$ and let $\rho_{XE}^{n}=\left(\cN_{AB}\otimes \id_E \right)\rho_{ABE}^{N}$ be a permutation-invariant state for a purification $\rho_{ABE}^N$ in $\cH_{ABE}^{\otimes N}$ of $\rho_{AB}^N \in \cS\left(\cH_{AB}^{\otimes N}\right)$. Then the rate of an $\eps{coh}:=(\eps{PE}+\eps{EC}+\eps{PA}+\epsbar)$-secure key is given by
\begin{equation}\label{eq:cohkey}
 r:= \frac{1}{N}\inf_{\rho^n_{AB}\in \Gamma^n_{\eps{PE}}}\left(\hmin{\epsbar}{\rho_{XE}^{n}}{E}-\leak \right)+\frac{2}{N}\log_2\left(2\eps{PA}\right).
\end{equation} 
\end{cor}
In the following section we show that the smooth min-entropy for permutation-invariant states can be mainly bounded by the min-entropy for corresponding product-states ``smoothed'' over a reduced $\eps{}$-environment.

\subsection{Privacy amplification}\label{subsec:PApermu}
In order to get a calculable formula for the key rate (Eq.~(\ref{eq:cohkey})) we bound the smooth min-entropy for permutation-invariant states by the smooth min-entropy for tensor-product states, which then can be easily evaluated by the asymptotic equipartition property (Eq.~(\ref{eq:minprod})) as explained in Section~\ref{SEC:Collective}.

We now define analogously to Eq.~(\ref{eq:permu}) the permutation-invariant state with $n$ signals, which Alice and Bob share after the parameter estimation procedure.
\begin{equation}\label{eq:rhosym}
 \rho_{AB}^n:=\cP_n\left(\sum_{\bn \in \Lambda^n} \mu_{\bn}\sigma_1^{\otimes n_1}\otimes \sigma_2^{\otimes n_2}\otimes \sigma_3^{\otimes n_3}\otimes \sigma_4^{\otimes n_4}\right),
 \end{equation} 
where $\sigma_i$ with $i=1,..,4$ correspond to the projectors onto the $4$ Bell-states in $\cH_A \otimes \cH_B$ and $\Lambda^n:=\left\lbrace \bn=\left(n_1,n_2,n_3,n_4\right):\sum_{i=1}^4 n_i =n \right\rbrace$ (see Eq.~(\ref{eq:permu})).
Additionally, we denote the single-copy state shared by Alice and Bob in the following as  
\begin{equation}
\sigma_{AB}[\blam]:=\sum_{i=1}^4 \lambda_i \sigma_i                      
\end{equation} with $\blam:=\left(\lambda_1,\lambda_2,\lambda_3,\lambda_4\right)=\left(\frac{n_1}{n},\frac{n_2}{n},\frac{n_3}{n},\frac{n_4}{n}\right)$.
 
The next theorem is one of our central results. It gives a relation between the smooth min-entropy for permutation-invariant states and the smooth min-entropy for tensor-product states. The proof is inspired by \cite{Ren05a} and uses the fact, that there exists a certain measurement on $\sigma_{AB}[\blam]^{\otimes n}$, such that the resulting state 
is equal to the state $\rho_{AB}^n$ for a specific realization $\bn$. Then, the application of some fundamental properties of the smooth min-entropy leads to the result.
\begin{thm} Let $\epsbar >0$, $\blam=\left(\frac{n_1}{n},\frac{n_2}{n},\frac{n_3}{n},\frac{n_4}{n}\right)$ and $\cM_{AB}$ be the quantum operation which describes the local measurements Alice and Bob perform followed by a partial-trace operation on Bob's part ($\cH_B$). Let $\rho_{XE}^n=\left(\cM_{AB} \otimes \id_E\right)^{\otimes n}\rho^n_{ABE}$  be the classical quantum state obtained after applying the quantum operation $\left(\cM_{AB} \otimes \id_E\right)^{\otimes n}$ on a purification $\rho_{ABE}^n$ in $\cH_{ABE}^{\otimes n}$ of a permutation-invariant state $\rho_{AB}^n \in \cS\left(\cH_{AB}^{\otimes n}\right)$. Analogously let $\sigma_{XE}[\blam]^{\otimes n}=\left(\cM_{AB} \otimes \id_E\right)^{\otimes n}\sigma_{ABE}[\blam]^{\otimes n}$ be the classical quantum state obtained after applying the quantum operation $\left(\cM_{AB} \otimes \id_E\right)^{\otimes n}$ on a purification $\sigma_{ABE}[\blam]^{\otimes n}$ of a tensor-product state $\sigma_{AB}[\blam]^{\otimes n} \in \cS\left(\cH_{AB}^{\otimes n}\right)$. Let $\cE$ be a $POVM$ on $\cH_A\otimes \cH_B$ which measures the $QBER$. Let $Q_n$, $P_n$ be an element of the frequency distribution $\bQ_n$, $\bP_n$ of the outcomes when applying the measurement $\cE^{\otimes n}$ to $\rho_{AB}^n$ and $\sigma_{AB}^{\otimes n}$, respectively. Then except with probability $\epsbar$
\begin{equation}\label{eq:PA}
\hmin{\epsbar}{\rho^n_{XE}}{E}\geq \hmin{\epsbar/(2n^2)}{\sigma_{XE}^{\otimes n}\left[\blam=\frac{\bn}{n}\right]}{E}-1,
\end{equation} 
where 
\begin{equation}\Gamma_{\mathrm{coh}}:=\left\lbrace \tau_{AB}
 : \dist{Q_m-P_n} \leq \xi_{\mathrm{coh}}\left(\epsbar,n,m\right)\right\rbrace
\end{equation} with 
\begin{equation}
\xi_{\mathrm{coh}}\left(\epsbar,n,m\right):=\frac{1}{2}\xi_{\mathrm{att}}\left(\epsbar,2,n\right)+\xi\left(\frac{\epsbar}{2},n,m\right) 
\end{equation} where
\begin{equation}
\xia(\epsbar,2,n):=\sqrt{\frac{16\ln{(2)}+8\ln{\left(1/\epsbar\right)}}{n}}
\end{equation} and
\begin{equation}
\xi\left(\frac{\epsbar}{2},n,m\right):=\sqrt{\frac{(m+n)(m+1)\ln{\left(2/\epsbar\right)}}{8m^2n}}
\end{equation} defines the set of tensor-product states $\tau^{\otimes n}$ which pass the parameter estimation procedure.
\end{thm}
\emph{Proof:} 
The state to be considered is given by $\rho_{XE}^n$ and can be expressed as a convex combination of states for all possible realizations $\bn$ with probability $\mu_{\bn}$, i.e.
\begin{equation}\label{eq:rhoXEsym}
 \rho_{XE}^n=\sum_{\bn \in \Lambda^n} \mu_{\bn}\rho_{XE}^n[\bn].
\end{equation}
 Note that this structure is provided in Eq.~(\ref{eq:rhosym}) and is conserved due to the linearity of $\cM_{AB}$ and a purification of $\rho_{AB}^n$, which is optimal for Eve.

 The first part proves the theorem for the special case, that only one $\mu_{\bn}$ in Eq.~(\ref{eq:rhoXEsym}) 
 is non-zero, i.e. we consider a single realization $\bn$. Then, part $2$ extends part $1$ to the general case.

\textbf{Part $1$:}

Let $\ket{\phi_i}$ be an extension to $\cH_A\otimes\cH_B\otimes\cH_E$ of $\sigma_i$ (see Eq.~(\ref{eq:rhosym})) with the condition, that the remaining states $\trz{AB}{P_{\ket{\phi_i}}}$ are mutually orthogonal for $i\in \left\lbrace 1,..,4\right\rbrace$. Note that this choice of orthogonal ancillas is optimal, since it enables the eavesdropper to distinguish perfectly the reduced states shared by Alice and Bob. Let $S_n$ be the set of distinguishable permutations $\pi$ on $n$ qubits for a fixed realization $\bn$. Then, with 
\begin{equation}
 \ket{\psi}^{\bn}_{ABE}:=\frac{1}{\sqrt{|S_n|}}\sum_{\pi \in S_n}\pi\left(\bigotimes_{i=1}^4 \ket{\phi_i}^{\otimes n_i}\right)
\end{equation}
 and 
\begin{equation}
 \ket{\phi}^{\blam}_{ABE}:=\sum_{i=1}^4 \sqrt{\lambda_i} \ket{\phi_i}
\end{equation} we define
\begin{eqnarray}
 \rho_{XE}^n[\bn]   &:=& \left(\cM_{AB} \otimes \id_E\right)^{\otimes n} P_{\ket{\psi}^{\bn}_{ABE}} \label{eq:rhoxe}\\
 \sigma_{XE}[\blam] &:=& \left(\cM_{AB} \otimes \id_E\right) P_{\ket{\phi}^{\blam}_{ABE}}
\end{eqnarray} for an arbitrary, but fixed realization $\bn$.
For any $i\in \left\lbrace 1,..,4\right\rbrace$ let $P_i$ be the projector onto the support of $\left(\cM \otimes \id_E\right) P_{\ket{\phi_i}}$ which by definition are orthogonal for distinct $i$. Let $\cF$ be a measurement defined by 
\begin{equation}
\cF: \rho \rightarrow \sum_{z=0}^1 F_z\rho F_z^\dagger \otimes \proj{z},
\end{equation} where 
\begin{equation}
 F_0:=\sum_{\pi \in S_n} \pi\left(P_1^{\otimes n_1}\otimes P_2^{\otimes n_2} \otimes P_3^{\otimes n_3} \otimes P_4^{\otimes n_4}\right)
\end{equation} and $F_1:=\id-F_0$. Then $F_0$ picks out a specific realization $\bn$ from the tensor-product state $\sigma_{XE}[\blam]^{\otimes n}$, i.e.
\begin{equation}\label{eq:meas}
 \rho^n_{XE}[\bn]=\frac{1}{P_Z(Z=0)}F_0\left(\sigma_{XE}[\blam]^{\otimes n}\right)F_0^\dagger
\end{equation} with $P_Z(Z=0)=\mathrm{tr}\left( F_0\left(\sigma_{XE}^{\otimes n}[\blam]\right)F_0^\dagger\right)=|S_n|\prod_{i=1}^4 \lambda_i^{n_i}$ (For a detailed proof see \cite{Ren05a}, Lemma~$A.4$).

Now let $\bar{\rho}^n_{XEZ}[\bn]$ be the resulting state after applying $\cF$ on $\sigma_{XE}^{\otimes n}[\blam]$ and let $Z$ be the classical measurement outcome, i.e.
\begin{eqnarray}
 \bar{\rho}^n_{XEZ}[\bn]
 &=& \sum_{z=0}^1 F_z \sigma_{XE}^{\otimes n}[\blam] F_z^\dagger \otimes \proj{z} \\
 &=:& \sum_{z=0}^1 P_Z(Z=z)\bar{\rho}_{XE}^{n Z=z}[\bn] \otimes \proj{z}.
\end{eqnarray}
 Then it follows directly from Eq.~(\ref{eq:meas}) that
\begin{equation}
 \rho^n_{XE}[\bn]=\bar{\rho}_{XE}^{n Z=0}[\bn]
\end{equation} and therefore
\begin{equation}\label{eq:condmeas}
 \hmin{\epsbar}{\rho^n_{XE}[\bn]}{E}=\hmin{\epsbar}{\bar{\rho}_{XE}^{n Z=0}[\bn]}{E}.
\end{equation} With some fundamental properties of the smooth min-entropy we get
\begin{eqnarray}\label{eq:part1a}
 && \hmin{\epsbar}{\bar{\rho}^{n Z=0}_{XE}[\bn]}{E} \nonumber \\
 &\stackrel{Eq.~(\ref{eq:cond})}{\geq}&\hmin{p_Z(Z=0)\epsbar}{\bar{\rho}^n_{XEZ}[\bn]}{EZ} \nonumber \\
 &\stackrel{Eq.~(\ref{eq:chain})}{\geq}&\hmin{p_Z(Z=0)\epsbar}{\bar{\rho}^n_{XEZ}[\bn]}{E}-\log_2{\left(\mathrm{rank}(\rho_Z)\right)}. \nonumber \\
 &&
\end{eqnarray}
By definition, the orthogonality and completeness of the set $\left\lbrace F_z\right\rbrace$ ensures that $\trz{Z}{\bar{\rho}^n_{XEZ}[\bn]}=\sigma_{XE}^{\otimes n}[\blam]$, such that we can apply Eq.~(\ref{eq:smoothmeas}) in the Appendix. This leads to
\begin{eqnarray}
 && \hmin{p_Z(Z=0)\epsbar}{\bar{\rho}^n_{XEZ}[\bn]}{E}-\log_2{\left(\mathrm{rank}(\rho_Z)\right)} \nonumber \\
 &\stackrel{Eq.~(\ref{eq:smoothmeas})}{\geq}& \hmin{p_Z(Z=0)\epsbar}{\sigma_{XE}^{\otimes n}[\blam]}{E}-\log_2{\left(\mathrm{rank}(\rho_Z)\right)} \nonumber \\
 &\geq& \hmin{\epsbar/n^2}{\sigma_{XE}^{\otimes n}[\blam]}{E}-1 \label{eq:part1},
\end{eqnarray} where we used in the last step that $\mathrm{rank}(\rho_Z)\leq 2$ and from Lemma~\ref{lem:multinom} in the Appendix that
\begin{equation}
 p_Z(Z=0)=|S_n|\prod_{i=1}^4 \lambda_i^{n_i} > 1/n^2.
\end{equation}
The following part generalizes the proof to the unrestricted case. 

\textbf{Part $2$:}

Now let $\rho_{ABE}^n:=P_{\ket{\psi}}$ with 
\begin{equation}
\ket{\psi}:=\sum_{\bn \in \Lambda^n}\sqrt{\mu_{\bn}}\ket{\psi}^{\bn}_{ABE}
\end{equation} be a purification of $\rho_{AB}^n$. For any $\bn \in \Lambda^n$ let $\cH^n_E$ be the smallest subspace of $\cH_E^{\otimes n}$ containing the support of the traces $\rho_E^n[\bn]=\trz{\cH_{AB}^{\otimes n}}{\rho_{ABE}^n[\bn]}$. By the definition of the vectors $\ket{\phi_i}$ as in part $1$, the subspaces $\cH_E^n$ are orthogonal for distinct $\bn \in \Lambda^n$. There exists a projective measurement $\cF'$ onto the subspaces $\cH_{AB}^{\otimes n}\otimes \cH_E^n$. Now let the state $\tilde{\rho}_{XEZ'}^n$ be the resulting state from the measurement $\cF'$ of the state $\rho_{XE}^n$ and let $Z' \in \Lambda^n$ be the classical outcome, i.e.
\begin{eqnarray}
\tilde{\rho}^n_{XEZ'}
&=& \sum_{\bn \in \Lambda^n}  F'_\bn \rho_{XE}^n F'^\dagger_\bn \otimes \proj{\bn} \\ \label{Eq.tilde}
&=:& \sum_{\bn \in \Lambda^n} \mu_\bn \rho_{XE}^n[\bn] \otimes \proj{\bn}.
\end{eqnarray} 
By the definition of the state $\rho_{XE}^n$ we know that for a tolerated $QBER$ $Q_m$ the parameter $Q_n$ for a virtual measurement on $n$ signals has to fulfil except with probability $\frac{\epsbar}{2}$ that
\begin{equation}\label{eq:PEhelp}
 \dist{Q_m-Q_n} \leq \xi\left(\frac{\epsbar}{2},n,m\right).
 \end{equation} 
Note that the choice of $\frac{\epsbar}{2}$ is arbitrary. In principle, the introduction of a new parameter could lead to better results. Now, this condition implies that realizations $\bn$ in the permutation-invariant state $\rho^n_{AB}=\sum_{\bn \in \Lambda^n} \mu_{\bn} \rho^n_{AB}[\bn]$, whose corresponding parameter $Q_n$ does not fulfill the condition in Eq.~(\ref{eq:PEhelp}),
only appear with small probability, i.e. more precisely
\begin{equation}
\sum_{\bn:\dist{Q_m-Q_n}>\xi(\frac{\epsbar}{2},n,m)} \mu_{\bn} \leq \frac{\epsbar}{2}.
\end{equation}
This behaviour of the probabilities enables us to apply Eq.~(\ref{eq:condclass}) in the Appendix for probability $\eps{}'=\frac{\epsbar}{2}$ to restrict the states $\rho_{AB}^n[\bn]$ (or equivalently their corresponding realizations $\bn$) to the set
\begin{equation}
 \tilde{\Gamma}^n_{\epsbar/2}:=\left\lbrace \sigma^n_{AB}[\bn]
:\dist{Q_m-Q_n} \leq \xi\left(\frac{\epsbar}{2},n,m\right)\right\rbrace.
\end{equation}
Namely, we have 
\begin{eqnarray}\label{EQ:help}
&& \hmin{\epsbar}{\rho^n_{XE}}{E} \nonumber \\
&\stackrel{Eq.~(\ref{eq:superpos})}{\geq}& \hmin{\epsbar}{\tilde{\rho}^n_{XEZ'}}{EZ'} \nonumber \\
&\stackrel{Eq.~(\ref{eq:condclass})}{\geq}& \inf_{\rho^n_{AB}[\bn]\in \tilde{\Gamma}^n_{\epsbar/2}} \hmin{\epsbar/2}{\rho^n_{XE}[\bn]}{E}. 
\end{eqnarray} 
Then Eq.~(\ref{EQ:help}) becomes, together with Eq.~(\ref{eq:condmeas}), Eq.~(\ref{eq:part1a}) and Eq.~(\ref{eq:part1}),
\begin{eqnarray}\label{EQ:help2}
&& \inf_{\rho^n_{AB}[\bn]\in \tilde{\Gamma}^n_{\epsbar/2}}\hmin{\epsbar/2}{\rho^n_{XE}[\bn]}{E} \nonumber \\
&\geq& \inf_{\rho^n_{AB}[\bn]\in \tilde{\Gamma}^n_{\epsbar/2}}\hmin{\epsbar/(2n^2)}{\sigma_{XE}^{\otimes n}[\blam=\frac{\bn}{n}]}{E}-1 \nonumber \\
&&
\end{eqnarray}
Since the min-entropy is now a function of a tensor-product state, we would like to express the restricting infimum in terms of the statistics $\bP_n$ of this tensor-product.
By definition, we have $\rho^1_{XE}[\bn]=\sigma_{XE}[\blam=\frac{\bn}{n}]$, such that we can apply Lemma~\ref{lem:chris} in the Appendix (for $k=N=n$), which states that, except with probability $\epsbar$, the statistics $\bP_n$ of the tensor-product state $\sigma_{XE}^{\otimes n}[\blam=\frac{\bn}{n}]$ is $\xi_{\mathrm{att}}$-close to $\bQ_n$, i.e.
\begin{equation}
 \dist{\bQ_n-\bP_n} \leq \xi_{\mathrm{att}}\left(\epsbar,|\cE|,n\right).
\end{equation} (Here the choice of $\epsbar$ is arbitrary. The consideration of a new parameter could in general lead to better results.) 
Now, we are able to bound the distance between $\bP_n$ and the tolerated $QBER$ $\bQ_m$ measured during parameter estimation by using the triangular inequality.
\begin{eqnarray}
 \dist{Q_m-P_n} 
 &\leq& \dist{Q_m-Q_n}+\dist{Q_n-P_n} \nonumber \\
 &\leq& \xi\left(\frac{\epsbar}{2},n,m\right)+\frac{\xia\left(\epsbar,2,n\right)}{2} \nonumber \\
 &=:& \xi_{\mathrm{coh}}\left(\epsbar,n,m\right),            
\end{eqnarray} where we used that for the $POVM$ applied for parameter estimation (see Eq.~(\ref{eq:PEcoll}) and Section~\ref{SUBSEC:PE}) the number of $POVM$ elements becomes $2$ (see \cite{Bra11}) and that \cite{Bra11}
\begin{equation}
 \dist{Q_n-P_n}\leq \frac{1}{2}\dist{\bQ_n-\bP_n}.
\end{equation} Consequently we end up in 
\begin{eqnarray}\label{EQ:help3}
 &&\inf_{\rho^n_{AB}[\bn]\in \tilde{\Gamma}^n_{\epsbar/2}} \hmin{\epsbar/(2n^2)}{\sigma_{XE}^{\otimes n}[\blam=\frac{\bn}{n}]}{E}-1 \nonumber \\
&\geq&\inf_{\sigma_{AB}\in\Gamma_{\xi_{\mathrm{coh}}}} \hmin{\epsbar/(2n^2)}{\sigma_{XE}^{\otimes n}\left[\blam=\frac{\bn}{n}\right]}{E}-1. \nonumber \\
&&
\end{eqnarray}
The assertion then follows by putting Eq.~(\ref{EQ:help3}) and Eq.~(\ref{EQ:help2}) into Eq.~(\ref{EQ:help}).\qed

Finally, we are able to formulate a calculable rate of an $\eps{coh}$-secure key for coherent attacks.
\begin{thm} Let $\eps{PE}, \eps{EC}, \eps{PA}, \epsbar > 0$ and let $\rho_{XE}^{n}=\left(\cN_{AB}\otimes \id_E \right)\rho_{ABE}^{N}$ be a permutation-invariant state for a purification $\rho_{ABE}^N$ in $\cH_{ABE}^{\otimes N}$ of $\rho_{AB}^N \in \cS\left(\cH_{AB}^{\otimes N}\right)$. Then the rate of an $\eps{coh}:=(\eps{PE}+\eps{EC}+\eps{PA}+2\epsbar)$-secure key is given by 
 \begin{eqnarray}\label{eq:ratecoh}
r_{\mathrm{coh}} &:=& \frac{n}{N}\Bigg[\inf_{\rho_{AB}\in\Gamma_{\mathrm{coh}}}\left( S(X|E)-\frac{\leak}{n} \right) \nonumber \\
                 &-& 5\sqrt{\frac{\log_2(4n^2/\epsbar)}{n}}\Bigg] \nonumber \\
                 &-& \frac{1}{N}+\frac{2}{N}\log_2{\left(2\eps{PA}\right)},
\end{eqnarray} where 
\begin{equation}
\Gamma_{\mathrm{coh}}=\left\lbrace \sigma_{AB}
:\dist{Q_m-P_n} \leq \xi_{\mathrm{coh}}\left(\epsbar,n,m\right)\right\rbrace 
\end{equation} with 
\begin{equation}\label{eq:xicoh}
 \xi_{\mathrm{coh}}\left(\epsbar,n,m\right):=\frac{\xia\left(\epsbar,2,n\right)}{2}+\xi\left(\frac{\epsbar}{2},n,m\right) 
\end{equation} for 
\begin{eqnarray}
\xi\left(\frac{\epsbar}{2},n,m\right)&:=&\sqrt{\frac{(m+n)(m+1)\ln{\left(2/\epsbar\right)}}{8m^2n}}, \\
\xia(\epsbar,2,n)&:=&\sqrt{\frac{16\ln{(2)}+8\ln{\left(1/\epsbar\right)}}{n}}\label{eq:xiatt}
\end{eqnarray} and 
\begin{equation}
S(X|E)=S(\rho_{XE})-S(\rho_E)
\end{equation} with $S(\rho):=-\tr{\left(\rho \log_2{\rho}\right)}$.
\end{thm}
\emph{Proof:}
 The proof follows by inserting the result from Eq.~(\ref{eq:PA}) 
 into Eq.~(\ref{eq:cohkey}) and using Eq.~(\ref{eq:minprod}) to express the smooth min-entropy of product states by the conditional von Neumann entropy of a single-copy state.\qed

A careful analysis of the proof of Eq.~(\ref{eq:PA}) enables us to obtain the main corrections for the secret key rate for coherent attacks (Eq.~(\ref{eq:ratecoh}))
 in comparison to collective attacks (Eq.~(\ref{eq:ratecoll})): First, for coherent attacks the probability to measure a single realization $\bn$ for a given tensor-product state is rather small, which makes the $\eps{}$-environment, e.g. in Eq.~(\ref{EQ:help3}) small. Second, the statistics for the different attacks are not identical in general. Additional fluctuations have to be taken into account as done by considering $\xi_{\mathrm{att}}$ (see Eqs.~(\ref{eq:xicoh}) and (\ref{eq:xiatt})).
These corrections loose their corrupting influence on the secret key rate, when considering the asymptotic 
limit ($N\rightarrow \infty$, $\eps{}\rightarrow 0$). In this case $\xi_{\mathrm{att}}$ becomes zero and no additional fluctuations have to be added to the $QBER$, thus the corrections vanish. This confirms the equivalence of collective and coherent attacks for permutation-invariant protocols stated in \cite{Kra05,Ren05a} in the asymptotic limit.
But for a finite number of signals these corrections have a dramatic impact on the secret key rate. And, since these additional terms seem
unavoidable, this might be a hint, that the equivalence of collective and coherent attacks might not hold for permutation-invariant states in the regime of finite resources. 

The following section shortly reviews the known post-selection technique \cite{Chri09}, which we then will compare to Eq.~(\ref{eq:ratecoh}).

\section{\label{SEC:PostSelection}Post-selection - A short review}
In order to determine the quality of $r_{\mathrm{coh}}$ (Eq.~(\ref{eq:ratecoh})) from the previous section, we have to compare it to key rates obtained by strategies existing in the literature. Up to now, there exist two main techniques to quantify secret key rates for finite resources for coherent attacks for the whole class of permutation-invariant protocols, namely the de Finetti approach \cite{RenPhD,Ren07} and the post-selection technique \cite{Chri09}. Since Sheridan et al showed in \cite{She10a} that the latter technique leads to higher secret key rates, we only take the post-selection technique for comparison.
 
The post-selection technique applied to QKD estimates the deviation of the finite key rate $r_{\mathrm{post}}$ obtained from a permutation-invariant protocol against coherent attacks from the corresponding rate $r_{\mathrm{coll}}$ against collective attacks. The rate of an $\eps{post}$-secure key is given by \cite{Chri09}
\begin{equation}\label{eq:ratecohPOST}
 r_{\mathrm{post}}=r_{\mathrm{coll}}-30\log_2{(N+1)}/N
\end{equation} where $r_{\mathrm{coll}}$ is given by Eq.~(\ref{eq:ratecoll}) evaluated for the security parameter $\eps{coll}=\eps{post}(N+1)^{-15}$.

\section{\label{SEC:Comparison}Comparison} 
In this section we compare our newly developed secret key rate $r_{\mathrm{coh}}$ (Eq.~(\ref{eq:ratecoh})) and the known rate $r_{\mathrm{post}}$ (Eq.~(\ref{eq:ratecohPOST})) for coherent attacks to the secret key rate evaluated under the assumption of collective attacks $r_{\mathrm{coll}}$ (Eq.~(\ref{eq:ratecoll})) for the BB$84$ protocol and the six-state protocol.
 
The finite-key rates are calculated for a total security parameter of $\eps{}:=\eps{coll}=\eps{post}=\eps{coh}=10^{-9}$. In the following let $QBER:=Q_m$ denote the tolerated $QBER$ from the POVM used for parameter estimation (see Eq.~(\ref{eq:PEcoll}) and Section~\ref{SUBSEC:PE}). 

The results are obtained from a numerical optimization procedure, which maximizes the key rate with respect to the parameters $m,\bar{\varepsilon},\eps{PE},\eps{EC},\eps{PA}$.

\begin{figure}[htbp]
    \centering
      \includegraphics[width=0.5\textwidth]{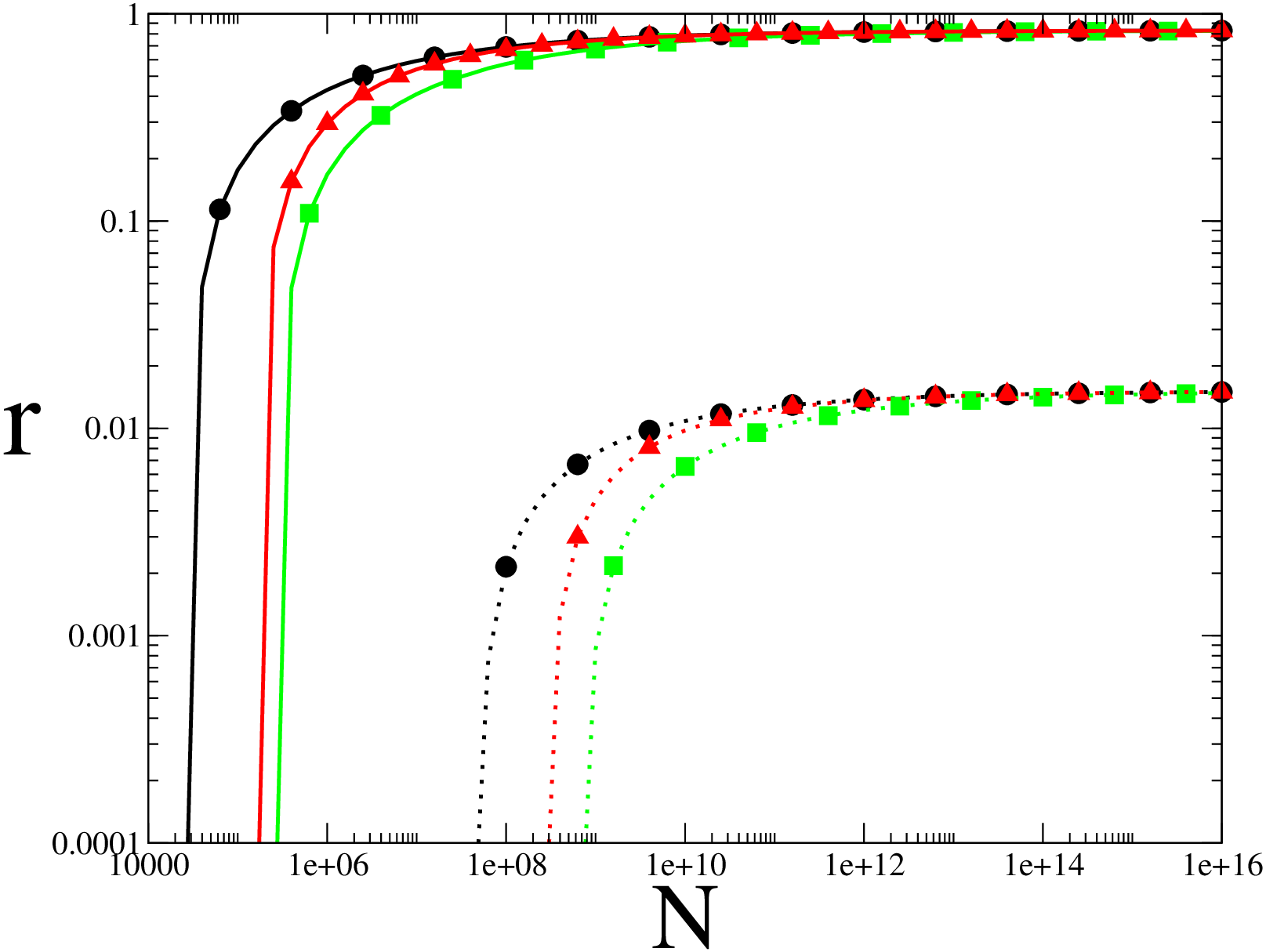}
    \caption{(Color online) Comparison of the secret key rates $r_{\mathrm{coll}}$ (Eq.~(\ref{eq:ratecoll})) (black circles), $r_{\mathrm{post}}$ (Eq.~(\ref{eq:ratecohPOST})) (green squares) and $r_{\mathrm{coh}}$ (Eq.~(\ref{eq:ratecoh})) (red triangles) versus the number $N$ of initial signals for different QBERs with security parameter $\eps{}=10^{-9}$ for the BB$84$ protocol in logarithmic scale; $QBER=0.01$ (straight lines), 
    $QBER=0.1$ (dotted lines).}
 \label{FIG:BB84}
\end{figure}

\begin{figure}[htbp]
    \centering
      \includegraphics[width=0.5\textwidth]{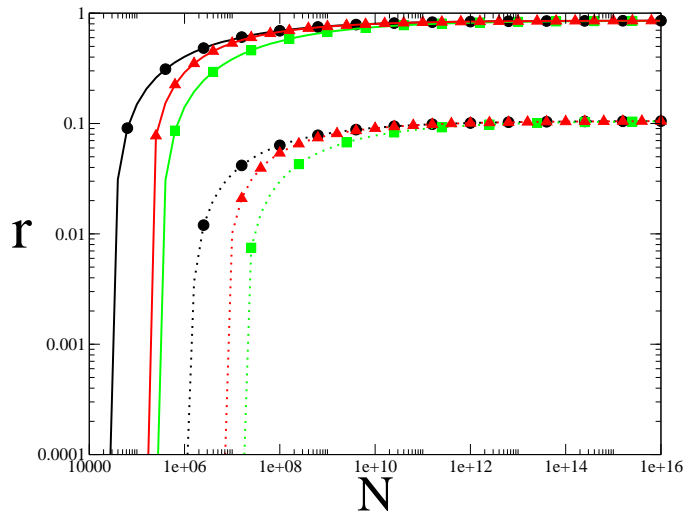}
    \caption{(Color online) Comparison of the secret key rates $r_{\mathrm{coll}}$ (Eq.~(\ref{eq:ratecoll})) (black circles), $r_{\mathrm{post}}$ (Eq.~(\ref{eq:ratecohPOST})) (green squares) and $r_{\mathrm{coh}}$ (Eq.~(\ref{eq:ratecoh})) (red triangles) versus the number $N$ of initial signals for different QBERs with security parameter $\eps{}=10^{-9}$ for the six-state protocol in logarithmic scale; $QBER=0.01$ (straight lines), 
    $QBER=0.1$ (dotted lines).}
 \label{FIG:sixstate}
\end{figure}

In FIG.~\ref{FIG:BB84} the secret key rates are shown as a function of the initial number of signals $N$ for different $QBER$s for the BB$84$ protocol.
FIG.~\ref{FIG:sixstate} presents an analogous calculation for the six-state protocol. 
Note that, as mentioned in Section~\ref{subsec:PApermu}, in both cases we recover the known result that coherent attacks become collective attacks in the limit of infinitely many signals $N$. For finite $N$ the figures show that the new rate $r_{\mathrm{coh}}$ is always significantly higher in comparison to the rate $r_{\mathrm{post}}$ obtained from the post-selection technique. This advantage of $r_{\mathrm{coh}}$ can be seen for a rather small $QBER=0.01$ as well as for a high value $QBER=0.1$. For example we obtain that the increase of $r_{\mathrm{coh}}$ in comparison to $r_{\mathrm{post}}$ is around $43\%$ for a $QBER$ of $0.01$ ($N=10^6$) and $33\%$ for a $QBER$ of $0.1$ ($N=10^{10}$) for the BB$84$ protocol. In case of the six-state protocol $r_{\mathrm{coh}}$ exceeds $r_{\mathrm{post}}$ by around $51\%$ for a $QBER$ of $0.01$ ($N=10^6$) and $45\%$ for a $QBER$ of $0.1$ ($N=10^{8}$).

\section{\label{SEC:Conclusion}Conclusion}
In this paper we presented a new method to quantify the rate of a secret key for general permutation-invariant protocols for coherent attacks. We show a technique to trace the calculation of secret key rates for coherent attacks back to the analysis of collective attacks. The high quality of this method manifests itself by a comparison to the up to now best-known strategy, 
 the post-selection technique. For the treatment of collective attacks we applied the von Neumann entropy bound. We showed that for a finite number of initial signals the secret key rates for the BB$84$ and the six-state protocol obtained by our method exceed the rates coming from the post-selection technique significantly. In case of the BB$84$ protocol, higher secret key rates have been obtained in \cite{Tom11} and \cite{Hay11} by a specialized method, which can, however, not be applied to the six-state protocol. Our method, in contrast, can be applied to all permutation-invariant quantum key distribution protocols for which an analysis of collective attacks is available. Since our results strongly depend on the underlying analysis of collective attacks, a prospective progress in the analysis of collective attacks will automatically cause a progress in our strategy with respect to secret key rates. 

Additionally the results of our derivation confirm the known result that, in the limit of infinitely many initial signals, coherent attacks are as powerful as collective 
attacks. Furthermore, we point out the main impact on the corrections for the key rate against coherent attacks in comparison to collective attacks. Since this 
extensive impact seems unavoidable, this might give some evidence for the inequivalence of the two types of attacks for finite resources.

Since the assumption of permutation-invariance is fairly weak (most protocols used in the literature are permutation-invariant or can be made to), the results of this paper can be widely applied. 
\begin{acknowledgments}
We would like to thank Silvestre Abruzzo, Renato Renner and Marco Tomamichel for helpful discussions. This work was financially supported by Deutsche Forschungsgemeinschaft (DFG) and Bundesministerium f\"ur Bildung und Forschung (BMBF), project QuOReP.
\end{acknowledgments}

\begin{appendix}
\section{}
\subsection{Properties of the (smooth) min-entropy}
\begin{lem}\label{lem:ptracesmooth}
Let $\rho_{ABZ}:=\sum_{z\in \cZ}P_Z(z)\rho_{AB}^z\otimes \proj{z}\in \cS\left(\cH_A\otimes\cH_B\otimes \cH_Z \right)$ be a classical-quantum state with $\rho_{AB}=\trz{Z}{\rho_{ABZ}}$ and $\sigma_B\in \cS\left(\cH_B\right)$, then
 \begin{equation}\label{eq:ptracesmooth}
  \hmin{\eps{}}{\rho_{ABZ}}{B} \geq \hmin{\eps{}}{\rho_{AB}}{B}.
 \end{equation}
\end{lem}
 \emph{Proof:} For any $\nu>0$ there exists $\bar{\rho}_{AB} \in \ball{\frac{\eps{}}{2}}{\rho_{AB}}$ such that for any $\sigma_B$
  \begin{equation*}
  \hmin{}{\bar{\rho}_{AB}}{\sigma_B}\geq \hmin{\eps{}}{\rho_{AB}}{\sigma_B}-\nu.
  \end{equation*} 
 Then it follows with Eq.~(\ref{eq:ptrace}) that
  \begin{eqnarray*}
  \hmin{}{\bar{\rho}_{ABZ}}{\sigma_B} \geq \hmin{}{\bar{\rho}_{AB}}{\sigma_B}.
  \end{eqnarray*} To conclude the proof it suffices to verify that $\bar{\rho}_{ABZ} \in \ball{\frac{\eps{}}{2}}{\rho_{ABZ}}$.
  \begin{eqnarray*}
  && \dist{\bar{\rho}_{ABZ}-\rho_{ABZ}} \leq \dist{\bar{\rho}_{AB}-\rho_{AB}}\leq \frac{\eps{}}{2},
  \end{eqnarray*} where we used the fact that the trace-distance cannot increase when applying a quantum operation (see \cite{RenPhD}, Lemma~$A.2.1$). The assertion then follows by choosing $\sigma_B$ such that
  \begin{equation*}
   \hmin{\eps{}}{\rho_{AB}}{\sigma_{B}}=\hmin{\eps{}}{\rho_{AB}}{B}
  \end{equation*} and the fact that
  \begin{equation*}
   \hmin{}{\bar{\rho}_{ABZ}}{B}\geq \hmin{}{\bar{\rho}_{ABZ}}{\sigma_{B}}.
  \end{equation*}\qed

\begin{lem}\label{lem:smoothmeas}
 Let $\rho_{AB}\in \cS\left(\cH_A\otimes\cH_B\right)$, $\left\lbrace \ket{z}\right\rbrace_z$ a family of orthogonal vectors in $\cH_Z$ and $\eps{}>0$. Then for a state $\bar{\rho}_{ABZ}:=\sum_{z\in\cZ}F_z\rho_{AB}F_z^\dagger\otimes \proj{z}$ with $\sum_{z\in \cZ}F_z^\dagger F_z=\id$ and $\trz{Z}{\bar{\rho}_{ABZ}}=\rho_{AB}$
  \begin{equation}\label{eq:smoothmeas}
   \hmin{\eps{}}{\rho_{AB}}{B}\leq \hmin{\eps{}}{\bar{\rho}_{ABZ}}{B}.
  \end{equation}
 \end{lem}
 \emph{Proof:} From the definition of $\bar{\rho}_{ABZ}$ it follows immediately that
  \begin{equation*}
  \hmin{\eps{}}{\trz{Z}{\bar{\rho}_{ABZ}}}{B}=\hmin{\eps{}}{\rho_{AB}}{B}.
  \end{equation*} Then the assertion follows with Lemma~\ref{lem:ptracesmooth}
\begin{equation}
\hmin{\eps{}}{\trz{Z}{\bar{\rho}_{ABZ}}}{B}\leq \hmin{\eps{}}{\bar{\rho}_{ABZ}}{B}.
\end{equation}\qed

\begin{lem}\label{lem:superpos}
 Let $\rho_{AB}\in \cS\left(\cH_A\otimes\cH_B\right)$, $\left\lbrace \ket{z}\right\rbrace_z$ a family of orthogonal vectors in $\cH_Z$ and $\eps{}>0$. Then for a state $\bar{\rho}_{ABZ}:=\sum_{z\in\cZ}P_Z(Z=z)F'_z\rho_{AB}F'^\dagger_z\otimes \proj{z}$ with $\sum_{z\in \cZ}F'^\dagger_z F'_z=\id$ and $\trz{Z}{\bar{\rho}_{ABZ}}=\rho_{AB}$
  \begin{equation}\label{eq:superpos}
   \hmin{\eps{}}{\rho_{AB}}{B}\geq \hmin{\eps{}}{\bar{\rho}_{ABZ}}{BZ}.
  \end{equation}
 \end{lem}
 \emph{Proof:} From the definition of $\bar{\rho}_{ABZ}$ it follows immediately that
  \begin{equation*}
  \hmin{\eps{}}{\trz{Z}{\bar{\rho}_{ABZ}}}{B}=\hmin{\eps{}}{\rho_{AB}}{B}.
  \end{equation*} Then the assertion follows from the strong subadditivity of the smooth min-entropy (see Eq.~(\ref{eq:strongsub})), i.e.
\begin{equation}
\hmin{\eps{}}{\trz{Z}{\bar{\rho}_{ABZ}}}{B}\geq \hmin{\eps{}}{\bar{\rho}_{ABZ}}{BZ}.
\end{equation}\qed

 \begin{lem}\label{lem:condclass}
 Let $\rho_{ABZ}=\sum_{z\in \cZ}P_{\mathrm{Z}}(z) \rho_{AB}^z \otimes \pr{z}$ be a classical quantum state and $\eps{}, \eps{}'> 0$, then 
 for any subset $\cZ' \subseteq \cZ$ such that $Prob[z\in \cZ']> 1-\eps{}'$,
  \begin{equation}\label{eq:condclass}
   \hmin{\eps{}+\eps{}'}{\rho_{ABZ}}{BZ}\geq \inf_{z \in \cZ'}\hmin{\eps{}}{\rho_{AB}^z}{B}.
  \end{equation}
 \end{lem}
 \emph{Proof:} For any $\nu>0$ and $z\in \cZ'$ there exists $\bar{\rho}_{AB}^z \in \ball{\frac{\eps{}}{2}}{\rho_{AB}^z}$ such that for any $\sigma_B^z$
  \begin{equation*}
  \hmin{}{\bar{\rho}^z_{AB}}{\sigma^z_B}\geq \hmin{\eps{}}{\rho^z_{AB}}{\sigma^z_B}-\nu.
  \end{equation*} Let
  \begin{equation*}
   \bar{\rho}_{ABZ}:=\sum_{z\in \cZ'}P_{\mathrm{Z'}}(z) \bar{\rho}_{AB}^z \otimes \pr{z}.
  \end{equation*}
 Then it follows with Eq.~(\ref{eq:cond1}) that
  \begin{eqnarray*}
  \hmin{}{\bar{\rho}_{ABZ}}{\sigma_{BZ}} &=& \inf_{z \in \cZ'}\hmin{}{\bar{\rho}_{AB}^z}{\sigma_B^z} \\
                                         &\geq&  \inf_{z \in \cZ'} \hmin{\eps{}}{\rho_{AB}^z}{\sigma_B^z}-\nu.
  \end{eqnarray*} To conclude the proof it suffices to verify that $\bar{\rho}_{ABZ} \in \ball{\frac{\eps{}+\eps{}'}{2}}{\rho_{ABZ}}$.
  \begin{eqnarray*}
  && \dist{\bar{\rho}_{ABZ}-\rho_{ABZ}} \\ 
  &\stackrel{Eq.~(\ref{eq:trace})}{=}& \sum_{z\in \cZ'}P_{\mathrm{Z'}}(z)\dist{\bar{\rho}^z_{AB}-\rho_{AB}^z} \\
  && + \sum_{z\in \cZ \setminus \cZ'}P_{\mathrm{Z\setminus Z'}}(z)\dist{\rho_{AB}^z} \\
  &\leq& \frac{\eps{}}{2} \sum_{z\in \cZ'} P_{\mathrm{Z'}}(z)+\frac{1}{2} \sum_{z\in \cZ \setminus \cZ'}P_{\mathrm{Z}\setminus \mathrm{Z'}}(z) \\
  &\leq& \frac{\eps{}+\eps{}'}{2}.   
  \end{eqnarray*} The assertion then follows by choosing $\sigma^z_{B}$ such that
  \begin{equation*}
   \hmin{\eps{}}{\rho^z_{AB}}{\sigma^z_{B}}=\hmin{\eps{}}{\rho^z_{AB}}{B}
  \end{equation*} and the fact that
  \begin{equation*}
   \hmin{}{\bar{\rho}_{ABZ}}{BZ}\geq \hmin{}{\bar{\rho}_{ABZ}}{\sigma_{BZ}}.
  \end{equation*}\qed

\begin{lem}\label{lem:cond}
 Let $\rho_{ABZ}=\sum_{z\in \cZ}P_{\mathrm{Z}}(z) \rho_{AB}^z \otimes \pr{z}$ be a classical quantum state and $\eps{z}:=P_{\mathrm{Z}}(z)\eps{}$, then
  \begin{equation}\label{eq:cond}
   \hmin{\eps{z}}{\rho_{ABZ}}{BZ}\leq \hmin{\eps{}}{\rho_{AB}^z}{B}.
  \end{equation}
 \end{lem}
 \emph{Proof:} For any $\nu>0$ and $z\in \cZ$ there exists $\rho'_{ABZ} \in \ball{\frac{\eps{z}}{2}}{\rho_{ABZ}}$ such that for any $\sigma_{BZ}$
  \begin{equation*}
  \hmin{}{\rho'_{ABZ}}{\sigma_{BZ}}\geq \hmin{\eps{z}}{\rho_{ABZ}}{\sigma_{BZ}}-\nu.
  \end{equation*} Then it follows with Eq.~(\ref{eq:incr1}) that
  \begin{equation*}
  \hmin{}{\rho'^z_{AB}}{\sigma^z_B} \geq \hmin{\eps{z}}{\rho_{ABZ}}{\sigma_{BZ}}-\nu.
  \end{equation*} To conclude the proof it suffices to verify that $\rho'^z_{AB} \in \ball{\frac{\eps{}}{2}}{\rho_{AB}^z}$.
  \begin{eqnarray*}
  \frac{\eps{z}}{2}&\geq& \dist{\rho'_{ABZ}-\rho_{ABZ}} \\
  &\stackrel{Eq.~(\ref{eq:trace})}{=}& \sum_{z\in \cZ}P_{\mathrm{Z}}(z)\dist{\rho'^z_{AB}-\rho_{AB}^z} \\
  &\geq& P_{\mathrm{Z}}(z)\dist{\rho'^z_{AB}-\rho_{AB}^z}.   
  \end{eqnarray*} The assertion then follows by choosing $\sigma_{BZ}$ such that 
  \begin{equation*}
   \hmin{\eps{z}}{\rho_{ABZ}}{\sigma_{BZ}}=\hmin{\eps{z}}{\rho_{ABZ}}{BZ}
  \end{equation*} and the fact that
  \begin{equation*}
   \hmin{}{\rho'^z_{AB}}{B} \geq \hmin{}{\rho'^z_{AB}}{\sigma^z_B}.
  \end{equation*}\qed

\subsection{Estimation of frequency distributions} 
\begin{lem}\label{lem:tom} Let $\eps{PE}>0$ and $0 \leq k\leq N$. Let $\rho^N \in \cS\left(\cH^{\otimes N}\right)$ be a permutation-invariant quantum state, and let $\cE$ be a $POVM$ on $\cH$ which measures the quantum bit error rate ($QBER$). Let $Q_k$ and $Q_{N-k}$ be the $QBER$s when applying the measurement $\cE^{\otimes k}$ and $\cE^{\otimes N-k}$, respectively, to different subsystems of $\rho^N$. Then except with probability $\eps{PE}$ it holds that
  \begin{equation}
   \dist{Q_{N-k}-Q_k}\leq  \xi(\eps{PE},N-k,k)
  \end{equation} with $\xi(\eps{PE},N-k,k):=\sqrt{\frac{N(k+1)\ln{\left(1/\eps{PE}\right)}}{8k^2(N-k)}}$.
  \end{lem}
  \emph{Proof:}
  It follows from the supplementary information (Note $2$) of \cite{Tom11} that with $\eps{PE}:=e^{-\frac{2k(N-k)}{N}\frac{k}{k+1}(2\xi(\eps{PE},N-k,k))^2}$
 \begin{equation}
  Prob[Q_n\geq Q_k+2\xi(\eps{PE},N-k,k)]\leq \eps{PE}.
 \end{equation} The assertion then follows by negation of the statement.\qed
\subsection{Multinomial distribution}
\begin{lem}\label{lem:multinom}
 Let $n \in \mathbb{N}$ and $\lambda_i=\frac{n_i}{n}$ for $i=1,..,4$ with $\sum_{i=1}^4n_i=n$. Then 
 \begin{equation}
  \frac{n!}{n_1!n_2!n_3!n_4!}\prod_{i=1}^4 \lambda_i^{n_i}>\frac{1}{n^2}
 \end{equation} for $n>500$.
\end{lem}
 \emph{Proof:} After applying the logarithm we get
  \begin{equation}
   \ln{\left(\frac{n!}{n_1!n_2!n_3!n_4!}\prod_{i=1}^4 \lambda_i^{n_i}\right)}=\ln{(n!)}-\sum_{i=1}^4 \ln{(n_i!)}+n_i \ln{\left(\frac{n_i}{n}\right)}.
  \end{equation}
 By using the Stirling-formula
 \begin{equation}
  \sqrt{2\pi n}\left(\frac{n}{e}\right)^n < n! < \left(1+\frac{1}{11n}\right) \sqrt{2\pi n}\left(\frac{n}{e}\right)^n
 \end{equation} we get for $n>0$
 \begin{eqnarray}
  && \ln{(n!)}-\sum_{i=1}^4 \ln{(n_i!)}+n_i \ln{\left(\frac{n_i}{n}\right)} \nonumber \\
  &>& \frac{1}{2}\ln{(2\pi n)}-\left(\sum_{i=1}^4 \frac{1}{2}\ln{(2\pi n_i)}+\ln{\left(1+\frac{1}{11n_i}\right)}\right) \nonumber \\
  &=& -\frac{3}{2}\ln{(2\pi n)}-\left(\sum_{i=1}^4 \frac{1}{2}\ln{\left(\frac{n_i}{n}\right)}+\ln{\left(1+\frac{1}{11n_i}\right)}\right) \nonumber \\
  &>& -\frac{3}{2}\ln{(2\pi n)}-4\ln{\left(\frac{12}{11}\right)},
 \end{eqnarray} where we used in the last line that $\frac{1}{2}\ln{\left(\frac{n_i}{n}\right)}<0$ and $\ln{\left(1+\frac{1}{11n_i}\right)}<\ln{\left(1+\frac{1}{11}\right)}$ for $n_i>0$ $\forall i=1,..,4$.
 After exponentiation we end up in
 \begin{equation}
  \frac{n!}{n_1!n_2!n_3!n_4!}\prod_{i=1}^4 \lambda_i^{n_i}>\frac{1}{(2\pi n)^{3/2}}\left(\frac{11}{12}\right)^4 > \frac{1}{n^2},
 \end{equation} which holds for $n>500$.\qed

\section{Known results}
Here, we review known results, which are crucial for derivations in the paper.

 \subsection{Properties of the (smooth) min-entropy}
 \begin{itemize}
  \item Chain rule (see \cite{RenPhD}, Theorem~$3.2.12$): Let $\rho_{ABC}\in \cS\left(\cH_A\otimes \cH_B \otimes \cH_C\right)$ and $\eps{}\geq 0$. Then for $\rho_C=\trz{AB}{\rho_{ABC}}$
                    \begin{equation}\label{eq:chain}
                    \hmin{\eps{}}{\rho_{ABC}}{B}\leq \hmin{\eps{}}{\rho_{ABC}}{BC}+\log_2{\left(\mathrm{rank}\left(\rho_C\right)\right)}.
                    \end{equation}
  \item Conditioning on classical information (see \cite{RenPhD}, Theorem~$3.2.12$): Let $\rho_{ABZ}:=\sum_{z\in \cZ}P_Z(z)\rho_{AB}^z\otimes \proj{z}\in \cS\left(\cH_A\otimes \cH_B\otimes \cH_Z\right)$ a classical-quantum state, then
                                               \begin{equation}\label{eq:cond1}
                                               \hmin{}{\rho_{ABZ}}{BZ}=\inf_{z\in\cZ}\hmin{}{\rho^z_{AB}}{B}.  
                                               \end{equation}
 \item Strong subadditivity (see \cite{RenPhD}. Theorem~$3.2.12$): Let $\rho_{ABC}\in \cS\left(\cH_A\otimes \cH_B \otimes \cH_C\right)$ and $\eps{}\geq 0$, then
       \begin{equation}\label{eq:strongsub}
        \hmin{\eps{}}{\rho_{ABC}}{BC}\leq \hmin{\eps{}}{\rho_{AB}}{B}.
       \end{equation}
 \item Partial-trace operation on classical subsystem can only decrease min-entropy (see \cite{RenPhD}, Lemma~$3.1.9$): Let $\rho_{ABZ}:=\sum_{z\in \cZ}P_Z(z)\rho_{AB}^z\otimes \proj{z}\in \cS\left(\cH_A\otimes \cH_B\otimes \cH_Z \right)$ be a classical-quantum state with $\rho_{AB}=\trz{Z}{\rho_{ABZ}}$ and $\sigma_B\in \cS\left(\cH_B\right)$, then
 \begin{equation}\label{eq:ptrace}
  \hmin{}{\rho_{ABZ}}{\sigma_B} \geq \hmin{}{\rho_{AB}}{\sigma_B}.
 \end{equation}
 \item  Quantum operations can only increase min-entropy (see \cite{Tom10}, Theorem~$18$): Let $\rho_{AB} \in \cS\left(\cH_A\otimes \cH_B \right)$ and let $\cE$ be a quantum operation such that $\bar{\rho}_{AC}=\left(\id_A\otimes \cE\right)\rho_{AB}$, then
                                                         \begin{equation}\label{eq:incr1}
                                                         \hmin{\eps{}}{\bar{\rho}_{AC}}{C} \geq \hmin{\eps{}}{\rho_{AB}}{B}. 
                                                         \end{equation}
  \item Trace-distance of mixtures (see \cite{RenPhD}, Lemma~$A.2.2$): Let $\rho_{AZ}:=\sum_{z\in \cZ}P_Z(z)\rho_A^z\otimes \proj{z} \in \cS\left(\cH_A\otimes \cH_Z\right)$ be a classical-quantum state and an analogous definition for $\rho'_{AZ}$, then
                                                             \begin{equation}\label{eq:trace}
                                                             \dist{\rho_{AZ}-\rho'_{AZ}}=\sum_{z\in \cZ}P_{\mathrm{Z}}(z)\dist{\rho_A^z-\rho'^z_A}. 
                                                             \end{equation}
  \item Smooth min-entropy of quantum tensor-product states (see \cite{RenPhD}, Corollary~$3.3.7$): Let $\rho^{\otimes n}_{XE}\in \cS\left(\left(\cH_X\otimes \cH_E\right)^{\otimes n}\right)$ a classical-quantum tensor-product state and $\eps{}\geq0$, then
                                                      \begin{equation}\label{eq:minprod}
                                                      \hmin{\eps{}}{\rho^{\otimes n}_{XE}}{E}\geq n\left(S(X|E)-5\sqrt{\frac{\log_2{\left(2/\eps{}\right)}}{n}}\right),
                                                      \end{equation} where $S(X|E)=S(\rho_{XE})-S(\rho_E)$ with $S(\rho):=-\tr{\left(\rho \log_2{\rho}\right)}$. 

 \end{itemize}

\subsection{Estimation of frequency distributions} 
   \begin{lem}\label{lem:chris}\cite{Chri04,Ren05a} Let $\eps{att}>0$ and $0 \leq k\leq N$. Let $\rho^N \in \cS\left(\cH^{\otimes N}\right)$ be a permutation-invariant quantum state, and let $\cE$ and $\cF$ be $POVMs$ on $\cH$ with $|\cE|$ and $|\cF|$ outcomes, respectively. Let $\bQ^{\cE}_k$ and $\bQ^{\cF}_{N-k}$ be the frequency distribution of the outcomes when applying the measurement $\cE^{\otimes k}$ and $\cF^{\otimes N-k}$, respectively, to different subsystems of $\rho^N$. Finally, let $\Omega$ be any convex set of density operators such that, for any operator $A$ on $n-1$ subsystems, the normalization of $\trz{n-1}{\id\otimes A \rho^n \id \otimes A^\dagger}$ is contained in $\Omega$.
Then except with probability $\eps{att}$, there exists a state $\sigma \in \Omega$ such that
   \begin{eqnarray}
   && \frac{k}{N}\dist{\bQ^{\cE}_k-\bP^{\cE}_k}+\frac{N-k}{N}\dist{\bQ^{\cF}_{N-k}-\bP^{\cF}_{N-k}} \nonumber \\
   && \leq \xia(\eps{att},|\cE|+|\cF|,N)
   \end{eqnarray} where $\bP^{\cE}_k$, $\bP^{\cF}_{N-k}$ denote the probability distributions of the outcomes when measuring $\sigma$ with respect to $\cE$ and $\cF$, respectively and $\xia(\eps{att},|\cE|+|\cF|,N):=\sqrt{\frac{8\ln{(2)}\left(|\cE|+|\cF|\right)+8\ln{\left(1/\eps{att}\right)}}{N}}$.
  \end{lem}

\end{appendix}

\end{document}